\documentclass[aps,prb,twocolumn,showpacs,superscriptaddress]{revtex4}  
\usepackage{graphicx}  
\usepackage{dcolumn}   
\usepackage{bm}        
\usepackage{amssymb}   

	\usepackage{amsmath}
	\usepackage{makeidx}
	\usepackage{amsfonts}
	\usepackage{subfigure}
	\usepackage{epsfig}
	\usepackage{float}
	\usepackage{pst-grad} 
	\usepackage{pst-plot} 
	\usepackage[colorlinks,hyperindex]{hyperref}
	\hypersetup
	{
		colorlinks,%
		citecolor=black,%
		linkcolor=black,%
		urlcolor=black,%
	}

\newcommand{\vect}[1]{\boldsymbol{#1}}

\makeindex
\begin{document}

	\author{Mohammad Reza Benam}
	\thanks{Present affiliation: Department of Physics, Payame Noor University, P.O. Box 19395-3697 Tehran, Iran}
	\affiliation{Department of Physics $\&$ Astronomy, University of British Columbia, Vancouver, British Columbia, Canada V6T 1Z1  }	
	\affiliation{Stewart Blusson Quantum Matter Institute, University of British Columbia, Vancouver, British Columbia, Canada V6T 1Z4}
	\author{Kateryna Foyevtsova}
	\thanks{foyevtsova@phas.ubc.ca}
	\author{Arash Khazraie}
	\author{Ilya Elfimov}
	\author{George A. Sawatzky}
	\affiliation{Department of Physics $\&$ Astronomy, University of British Columbia, Vancouver, British Columbia, Canada V6T 1Z1  }	
	\affiliation{Stewart Blusson Quantum Matter Institute, University of British Columbia, Vancouver, British Columbia, Canada V6T 1Z4}

\title{Holes' character and bond versus charge disproportionation
in $s\textendash p$ $ABX_{3}$ perovskites}


\begin{abstract}
We use density functional theory methods to study the electronic structures
of a series of $s\textendash p$ cubic perovskites $ABX_{3}$: the experimentally
available SrBiO$_{3}$, BaBiO$_{3}$, BaSbO$_3$, CsTlF$_{3}$, and CsTlCl$_{3}$,
as well as the hypothetical MgPO$_{3}$, CaAsO$_{3}$, SrSbO$_{3}$,
and RaMcO$_3$.
We use tight-binding modeling to calculate the
interatomic hopping integrals $t_{sp\sigma}$ between
the $B\textendash s$ and $X\textendash p$
atomic orbitals and
charge-transfer energies $\Delta$, which are the two most important parameters
that determine the low-energy electron and hole states
of these systems. Our calculations elucidate
several trends in $t_{sp\sigma}$ and $\Delta$ as one moves
across the periodic table, such as the relativistic energy lowering
of the $B\textendash s$ orbital in heavy $B$ cations leading to strongly negative
$\Delta$ values.
Our results are discussed in connection with the general phase
diagram for $s\textendash p$ cubic perovskites proposed
in Ref.~\onlinecite{Khazraie2}, where the parent superconductors
SrBiO$_{3}$ and BaBiO$_{3}$ are found to be in the regime
of negative $\Delta$ and large $t_{sp\sigma}$.
Here, we explore this further and search for new materials
with similar parameters, which could lead to the discovery
of new superconductors.
Also, some considerations are offered regarding a possible relation between
the physical properties of a given $s\textendash p$ compound
(such as its tendency to bond disproportionate and
the maximal achievable superconducting transition temperature)
and its electronic structure.

\end{abstract}

\maketitle

\section{Introduction}
Materials with a cubic perovskite structure $ABX_{3}$,
where the anion $X$ can be an oxygen or a halogen and the
possible cations $A$ and $B$ include a broad variety of elements
or even molecules, have attracted considerable attention due
to their rich physics. Indeed, among their intriguing properties
are metal-insulator transitions\cite{Medarde, Imada},
high transition
temperature ($T_{\text c}$) superconductivity\cite{Bendorz, Sleight, Cava},
ferroelectricity, ferromagnetism, applicability in photovoltaics\cite{Green},
colossal magnetoresistance \cite{Ramirez},
magnetoelectricity \cite{Kimura}, and a topological insulating
state\cite{Jin, Jin2, Yan}.
The crystal structure of the
$ABX_{3}$ cubic perovskites consists of a three-dimensional
network of corner-sharing $BX_6$ octahedra intercalated with $A$
cations at the twelve-fold anion-coordinated sites.
One of the well-known and widely studied $ABX_{3}$ compounds is $A$BiO$_{3}$,
with $A$ = Ba or Sr. Upon hole doping, achieved via chemical substitutions,
these systems become superconducting with a surprisingly high maximal
$T_{\text c}$ of 30~K\cite{Sleight, Cava, Mattheiss, Kazakov}.
As stoichiometry is approached in the pure parent compound, however,
the superconductivity gives way
to an insulating state featuring a so-called breathing structural distortion,
where the BiO$_{6}$ octahedra disproportionate into small and large ones in
a rock-salt-like pattern\cite{Cox, Cox2, Zhou, Sleight2}.

\begin{figure}[tb]
\begin{center}
\includegraphics[width=0.49\textwidth]{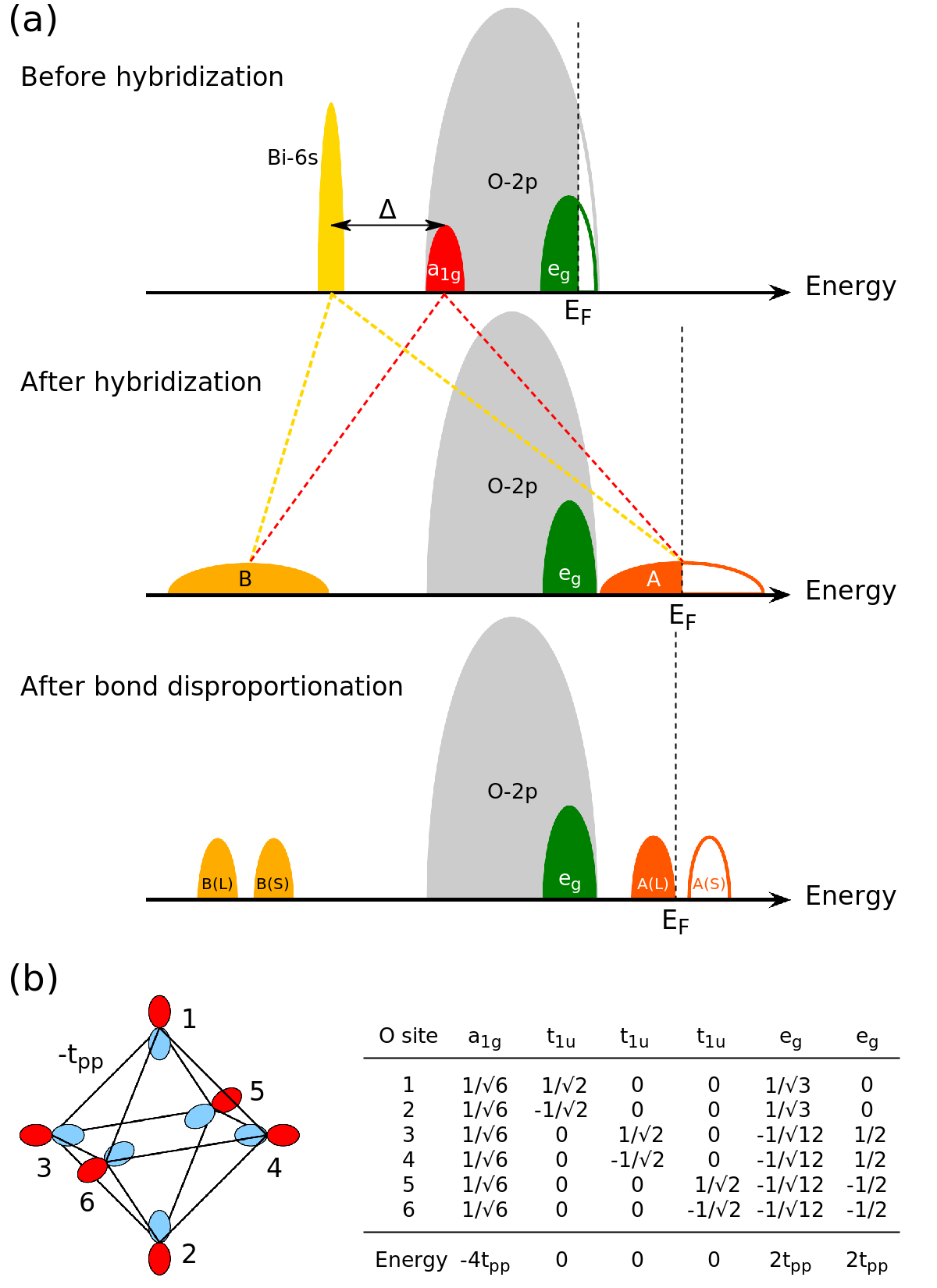} 
\caption{(a)~A schematic diagram of the Bi$\textendash 6s$
and O$\textendash 2p$ energy levels
in $A$BiO$_{3}$ before (top panel) and after (middle panel) hybridization.
''A`` and ''B`` denote an anti-bonding and a bonding band, respectively.
The bottom panel demonstrates the effect of Bi-O bond disproportionation, whereby
the bonding and the anti-bonding bands are each split into two
subbands associated with large and small BiO$_6$
octahedra, denoted as ''B(L)``, ''B(S)``, ''A(L)``, and
''A(S)``, respectively, and a charge gap is opened as a result.
(b)~The six O$\textendash 2p_{\sigma}$ orbitals of an O$_{6}$ octahedron and their
molecular orbital combinations, $a_{1g}$, $t_{1u}$,
and $e_g$, with their corresponding
energies in units of nearest-neighbor $pp\sigma$ hopping integral
$-t_{pp}$.
}
\label{fig1}
\end{center}
\end{figure}

Although in the early years following the discovery of $A$BiO$_{3}$,
their breathing distortion was viewed as a result of charge disproportionation
of the nominally tetravalent Bi$^{4+}$ ions into
Bi$^{3+}$ and Bi$^{5+}$\cite{Varma, Taraphder, Hase},
recent theoretical\cite{Mattheiss2, Harrison, Foyevtsova, Khazraie, Dalpian, Khazraie2}
as well as experimental\cite{Hair, Orchard, Wertheim, Salem, Plumb, Balandeh}
studies have seriously challenged this idea.
In a more realistic microscopic picture,
developed by some of us in Refs.~\onlinecite{Foyevtsova, Khazraie, Khazraie2},
one starts by recognizing the negative charge-transfer nature of
the $A$BiO$_{3}$ electronic states, {\it i.~e.},
that the O$\textendash 2p$ states are in fact higher in energy than the semi-core
Bi$\textendash 6s$ states (by amount $\Delta$),
as depicted in the top panel of Fig.~\ref{fig1}~(a).
However, the most important parameter of all shaping
the $A$BiO$_{3}$ electronic structure
is the strong hybridization between the Bi$\textendash 6s$ atomic orbital
and the $a_{1g}$ molecular orbital (MO) formed by the O$\textendash 2p_{\sigma}$ orbitals
of the oxygen octahedral cage [see Fig.~\ref{fig1}~(b)].
It produces a huge splitting between the bonding and anti-bonding bands,
much larger than the charge transfer energy,
with the latter band landing at the Fermi energy above the O non-bonding states,
as depicted in the middle panel of Fig.~\ref{fig1}~(a).
Since the character of the conductance anti-bonding band is predominantly that
of the O$\textendash a_{1g}$ MO, the average Bi oxidation state approaches 3+,
leaving two self-doped ligand holes, $\underline{L}$, per oxygen octahedron
as $2{\text{Bi}}^{4+} \rightarrow 2{\text{Bi}}^{3+}\underline{L}^{2}$,
in what Alex Zunger and co-workers called
a ``self-regulating response"\cite{Dalpian, Raebiger}.
Upon the breathing distortion,
resulting from the strong electron - breathing phonon
interaction, which increases the short-bond length
Bi$\textendash 6s-$O$\textendash 2p$ hopping integrals and therefore
stabilizes further the bonding state,
the ligand holes condense pairwise onto
the small octahedra as
$2{\text{Bi}}^{3+}\underline{L}^{2} \rightarrow
[{\text{Bi}}^{3+}]_{\text{large}}
+
[{\text{Bi}}^{3+}\underline{L}^{2}]_{\text{small}}$,
resulting in nearly the same valence states for the two inequivalent
Bi ions\cite{Foyevtsova},
a situation to be called bond, rather than charge,
disproportionation. As shown in the bottom
panel of Fig.~\ref{fig1}~(a), this process is associated with
opening of a charge gap at the center of the anti-bonding band
of the cubic structure.

With this picture of the $A$BiO$_{3}$ electronic structure in mind,
a further step was taken in Ref.~\onlinecite{Khazraie2} and a general phase diagram
was proposed to describe a crossover from a bond- to a charge-disproportionated
regime in $s\textendash p$ cubic perovskites similar to $A$BiO$_{3}$.
It was shown that there are two main electronic parameters that determine
the regime a given system will end up in: the charge-transfer energy $\Delta$
and the hybridization between the $B$-cation $s$ orbital and the oxygen
$a_{1g}$ MO characterized by the hopping integral $T_{sp\sigma}$.
However, even though a number of $s\textendash p$ cubic perovskites
other than $A$BiO$_{3}$ are known,
such as the recently synthesized CsTlCl$_{3}$ and CsTlF$_{3}$\cite{Retuerto},
and some of them even superconduct (BaPb$_{1-x}$Sb$_{x}$O$_{3}$\cite{Singh,Cava2}),
no real examples were discussed in Ref.~\onlinecite{Khazraie2}
in relation with the proposed
phase diagram. In order to fill this gap, in the present paper
we use {\it ab initio} theoretical methods to study the electronic structures
of the above mentioned existing $s\textendash p$ cubic perovskites
and also of a systematic
series of hypothetical $AB$O$_{3}$ systems,
with $A$ and $B$ cations being
the group IIa and group Va elements, respectively.
We hope that this study, conducted in the light of the notion of charge
{\it versus}
bond disproportionation, will add to our understanding of superconductivity
in the $s\textendash p$ cubic perovskites and also guide the discovery of new superconductors.

\section{Method}
Our electronic structure calculations are performed within
density functional theory (DFT)\cite{Kohn} using
the full-potential linearized augmented plane-wave 
method as implemented in the WIEN2k package\cite{Blaha}.
We employ the generalized gradient approximation (GGA)\cite{Perdew}
for the exchange-correlation potential.
For all our $s\textendash p$ systems,
a simplified cubic $Pm\bar{3}m$ crystal structure is assumed,
with both the tilting and breathing distortions neglected,
and the volume is fully relaxed within GGA.
The basis set size is fixed by setting $R_{\text{MT}}K_{\text{max}}=7$,
where $R_{\text{MT}}$ is the smallest muffin-tin sphere radius and $K_{\text{max}}$
is the cut-off wave vector. A $12\times12\times12$
grid of $\vect{k}$-points is used for integrating
over the first Brillouin zone.
Atomic and molecular orbital projections are done within muffin-tin spheres.
Projections onto molecular orbitals are done with a modified
version of WIEN2k, as discussed in Ref.~\onlinecite{Foyevtsova2}.
Tight-binding (TB) parameters are obtained by using
the maximally localized Wannier functions (MLWF) method
as implemented in the wannier90 code\cite{Mostofi}.

\begin{figure*}[tb]
\begin{center}
\includegraphics[width=0.98\textwidth]{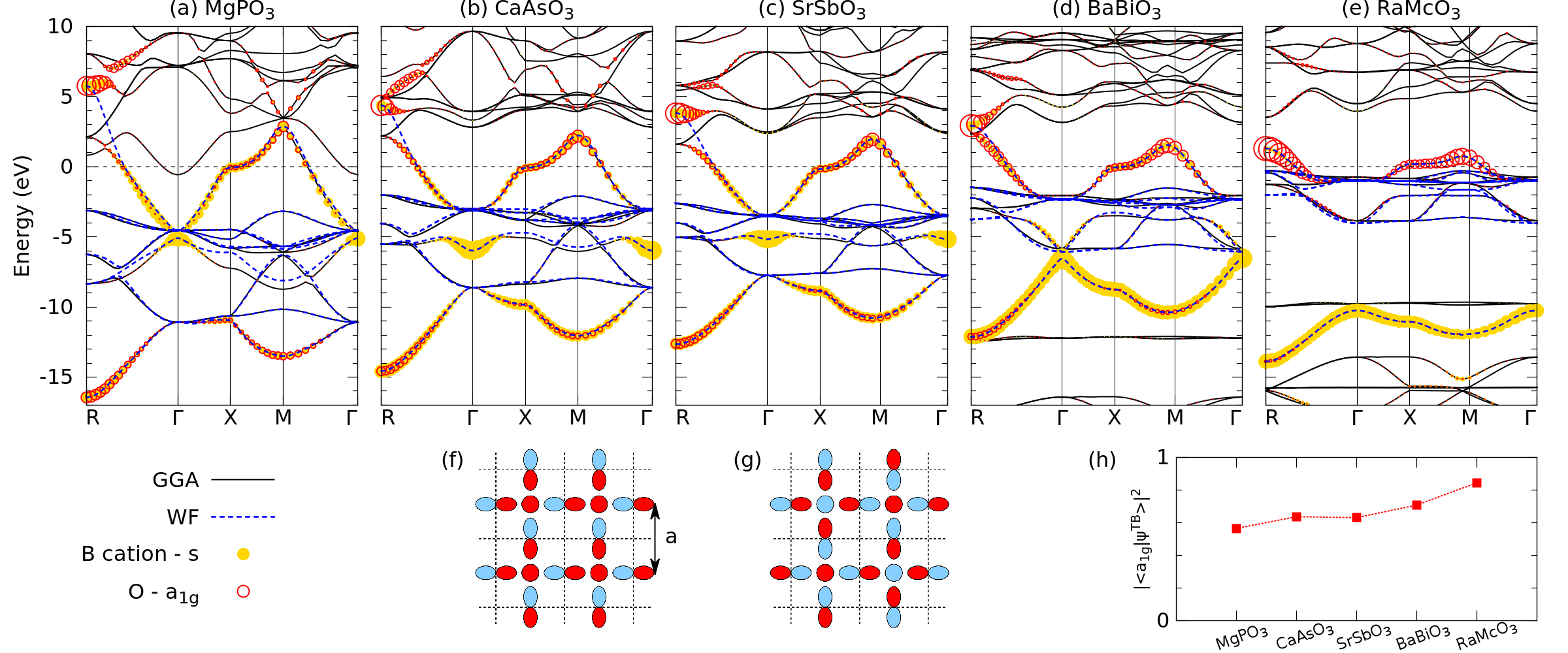} 
\caption{The electronic band-structures of 
(a)~MgPO$_{3}$,
(b)~CaAsO$_{3}$,
(c)~SrSbO$_{3}$,
(d)~BaBiO$_{3}$,
and (e)~RaMcO$_{3}$
calculated in GGA and plotted with black solid lines.
The Fermi energy is set to zero and marked by a horizontal black dashed line.
The red and yellow circles indicate the presence of, respectively,
the O$\textendash a_{1g}$ and cation $B\textendash s$ orbital characters
in a given Bloch eigenstate,
with the amount of their contribution being proportional to the circles' radii.
The dashed blue lines represent the eigenstates of the MLWF-based TB model.
Panels (f) and (g) use a two-dimensional (2D) analogue
of the cubic perovskite structure to explain the absence of hybridization
between the O$\textendash 2p_{\sigma}$ and $B\textendash s$ Bloch functions
at the $\Gamma$ point [$\vect{k} = (0, 0, 0)$]
and its maximum strength at the $R$ point [$\vect{k} = (\pi,  \pi,  \pi)$], respectively.
The dashed lines mark boundaries between 2D unit cells with the lattice constant $a$.
(h) The amount of the O$\textendash a_{1g}$
molecular orbital character in the anti-bonding band
of the $AB$O$_3$ TB models,
$|\langle a_{1g}(\vect{k}) | \psi^{\text{TB}}(\vect{k})\rangle|^2$,
at the $R$ point
as a function of
chemical composition.}
\label{fig2}
\end{center}
\end{figure*}

\begin{figure}
\begin{center}
\includegraphics[width=0.49\textwidth]{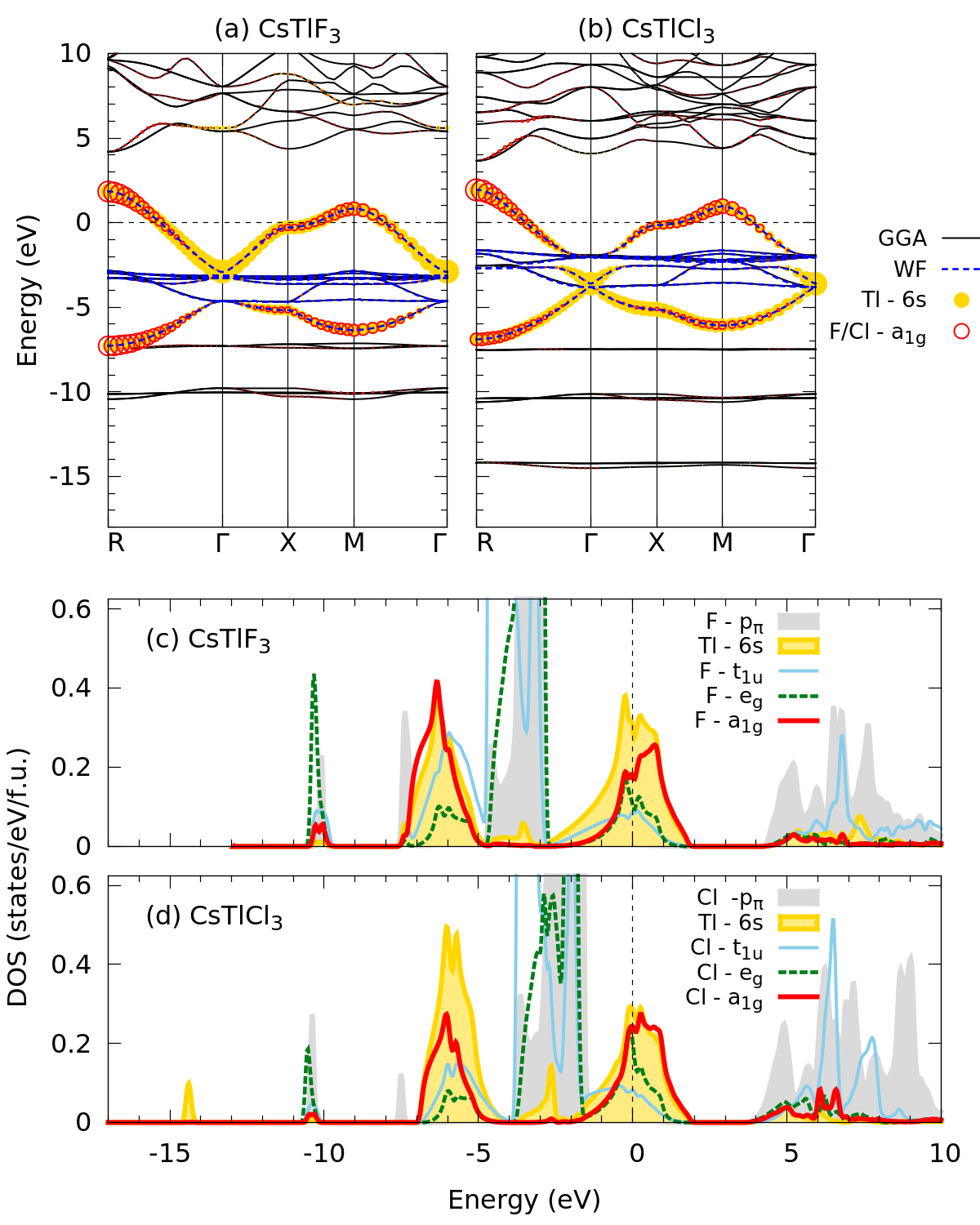} 
\caption{(a), (b) The electronic band-structures
of CsTlF$_{3}$ and CsTlCl$_{3}$.
Notations are similar to those in Fig.~\ref{fig2}.
(c), (d) Partial densities
of states of CsTlF$_{3}$ and CsTlCl$_{3}$,
projected onto the Tl$\textendash 6s$
atomic orbital and the molecular combinations
of the F$\textendash 2p_{\sigma}$ or Cl$\textendash 3p_{\sigma}$
orbitals. The zero of energy is at the Fermi energy.
}
\label{fig4}
\end{center}
\end{figure}

\begin{figure}[tb]
\begin{center}
\includegraphics[width=0.49\textwidth]{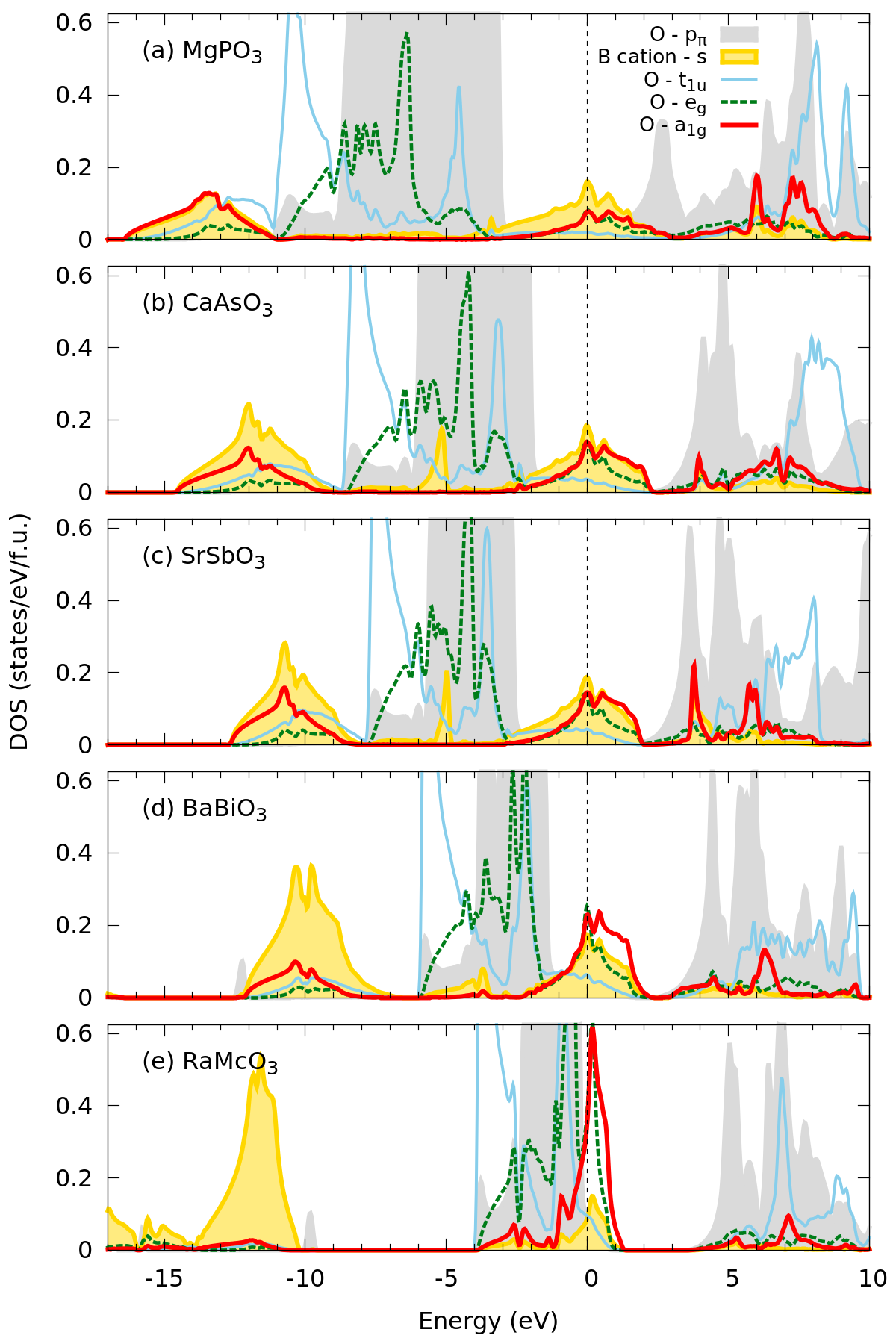} 
\caption{The partial densities of state projected
onto the $B\textendash s$ atomic orbital and the molecular
combinations of O$\textendash 2p_{\sigma}$ orbitals of (a)~MgPO$_{3}$,
(b)~CaAsO$_{3}$, (c)~SrSbO$_{3}$, (d)~BaBiO$_{3}$ and (e)~RaMcO$_{3}$.
The Fermi energy is set to zero and marked by a vertical black solid line.}
\label{fig3}
\end{center}
\end{figure}

\begin{table}[tb]
\caption{GGA equilibrium lattice constants $a$,
$B\textendash s - X\textendash p$ hopping integrals $t_{sp\sigma}$,
and charge-transfer energies $\Delta$ of the studied $ABX_{3}$ cubic perovskites.}
\begin{tabular}{l c c c c}
\hline\hline
&Has~been synthesized?& $a$~(\AA) &  $t_{sp\sigma}$~(eV)   & $\Delta$~(eV)  \\
\hline
MgPO$_{3}$    &No & 3.667                 & 2.65 & 1.83   \\
CaAsO$_{3}$   &No & 3.919                 & 2.30 & -1.44   \\
SrSbO$_{3}$   &No & 4.233                 & 2.08 & -1.08   \\
BaSbO$_{3}$   &Yes\cite{Cava2}   & 4.280  & 2.01 & -1.53   \\
SrBiO$_{3}$   &Yes\cite{Kazakov} & 4.372  & 1.88 & -3.68  \\
BaBiO$_{3}$   &Yes\cite{Sleight} & 4.417  & 1.81 & -3.96  \\
RaMcO$_{3}$   &No & 4.676                 & 1.46 & -8.86  \\
\\
CsTlF$_{3}$   &Yes\cite{Retuerto}& 4.799  & 1.24 & 0.65   \\
CsTlCl$_{3}$  &Yes\cite{Retuerto}& 5.604  & 1.16 & -0.88   \\
\hline\hline
\end{tabular} 
\label{tab1}
\end{table}

\section{Results and discussion}
Let us first discuss the systematic series of $AB$O$_{3}$
$s\textendash p$ cubic perovskites,
where the $A$ and $B$ cations are varied down the periodic table as MgPO$_{3}$,
CaAsO$_{3}$, SrSbO$_{3}$, BaSbO$_3$, SrBiO$_{3}$, BaBiO$_{3}$, and RaMcO$_{3}$.
Among them, only SrBiO$_{3}$, BaSbO$_3$, and BaBiO$_{3}$ exist in nature
(see Table~\ref{tab1}),
but our prime interest is to identify general trends in the electronic
structure of such $s\textendash p$ systems.

Figures~\ref{fig2}~(a)-(e) and \ref{fig3}~(a)-(e) show the band-structures
and the projected densities of states (DOS)
of the studied $AB$O$_3$ series, respectively.
Note that no results for SrBiO$_{3}$ and BaSbO$_3$ are shown as they are very
similar to those for, respectively, BaBiO$_{3}$ and SrSbO$_{3}$.
All the systems demonstrate
the same strong bonding\textendash anti-bonding splitting between the
$B\textendash s$ atomic and O$\textendash a_{1g}$ molecular orbitals, 
with the anti-bonding band landing at the Fermi level and becoming half occupied.
Only in MgPO$_{3}$,
there is a second band crossing the Fermi level,
which is mainly of the Mg$\textendash 3s$ orbital character.
As a general trend, the overall band width of
the $B\textendash s - {\text O}\textendash 2p$ states
decreases as we go down from MgPO$_{3}$ to RaMcO$_{3}$.
Indeed, as the lattice constant increases
due to the large ionic radii of the $A$ and $B$ cations
(see Table~\ref{tab1}),
the direct hopping between oxygen orbitals is decreasing,
which results in reduction of the oxygen orbitals' band width,
and so does the hopping between oxygen orbitals and the $B$ cation
$s$ orbitals, which results in reduction of the band width
of the $B\textendash s$ and O$\textendash a_{1g}$ hybrid.
On one hand, the narrowing of the oxygen band makes it easier
to push the $a_{1g}$ states up and out of the top of the
oxygen band. On the other hand,
the reduced $B\textendash s - {\text O}\textendash 2p_{\sigma}$
hopping integral $t_{sp\sigma}$
leads to a flattening of the anti-bonding conduction band
in BaBiO$_3$ and, especially, RaMcO$_3$.
Correspondingly, the density of states
at the Fermi level is strongly increased in these two end members
of the series, which makes them more strongly driven
towards bond disproportionation and
other types of structural distortions.
Another important trend
in the band structures of the $AB$O$_3$ series
is
the gradual change of the dominating character of the
anti-bonding conduction band from one of more $B\textendash s$ atomic orbital
to one of more O$\textendash a_{1g}$ MO character.
This is illustrated in
Fig.~\ref{fig2}~(h) showing
the amount of the O$\textendash a_{1g}$
MO character in the anti-bonding band
of the $AB$O$_3$ TB models (which will be discussed in more detail shortly),
$|\langle a_{1g}(\vect{k}) | \psi^{\text{TB}}(\vect{k})\rangle|^2$,
at the $R$ point
as a function of
chemical composition.

It is important to note that the bonding\textendash anti-bonding splitting is
strongly $\vect{k}$-vector dependent,
due to the changing symmetry of the Bloch functions involved.
The splitting
vanishes at $\Gamma$ [$\vect{k} = (0, 0, 0)$], because
at this point the Bloch
wave function has no $a_{1g}$
molecular orbital component [Fig.~\ref{fig2}~(f)].
In contrast, at the $R$ point [$\vect{k} = (\pi, \pi, \pi)$],
the Bloch wave function is of a pure
$a_{1g}$
molecular orbital character in the oxygen $p_{\sigma}$ orbitals'
domain [Fig.~\ref{fig2}~(g)], and the splitting
reaches its maximum.

In accordance with this logic, the
band structure plots of Fig.~\ref{fig2}
feature no
$a_{1g}$ character in any of the Bloch states at the $\Gamma$ point.
Also, there is only one state with dominating $B\textendash s$
character at this $\vect{k}$-vector, which moves to
lower energies towards the end of the $AB$O$_3$ series,
from around -5~eV in MgPO$_3$ to -10~eV in RaMcO$_3$.
Along the $\Gamma-R$ path (and, actually,
at any point other than $\Gamma$), 
the finite $B\textendash s - {\text O}\textendash 2p_{\sigma}$
hybridization is expected to make the bonding band disperse downwards
and reach a minimum at $R$.
While this is what we indeed observe
in the band structures of BaBiO$_3$ and RaMcO$_3$, the
behavior of the bonding band in the band structures of
MgPO$_3$, CaAsO$_3$ and SrSbO$_3$ appears to
be more complex. There, the $B\textendash s$ state
at $\Gamma$
happens to be energetically above the triply degenerate
oxygen molecular orbital $t_{1u}$ states
(positioned at around -11~eV in MgPO$_3$
and -8~eV in CaAsO$_3$ and SrSbO$_3$). Away from $\Gamma$,
these oxygen bands disperse upwards and get
entangled with the bonding band. This is the reason
why in MgPO$_3$, CaAsO$_3$ and SrSbO$_3$
the character of the bonding $a_{1g}$ and
$B\textendash s$ combination is not continuous.

A similar situation occurs for
the anti-bonding band close to the $R$ point.
At this point, in the same three systems, the interatomic
orbital hybridizations
are strong enough to push the anti-bonding
state above the triply degenerate $B\textendash p$ states
and position it at around 4 to 6~eV.
Away from $R$, the $B\textendash p$ bands
and the anti-bonding band mix with each other,
and this is the reason why in these three systems
the character of the anti-bonding $a_{1g}$ and
$B\textendash s$ combination
is not
continuous either.

Let us now
compare our findings about the $AB$O$_3$ series with the calculated electronic
structures of CsTlF$_{3}$ and CsTlCl$_{3}$, shown in Fig.~\ref{fig4}.
These recently synthesized halides\cite{Retuerto}
also demonstrate the bond disproportionation,
and so qualitatively
we expect their electronic structure near $E_{\text F}$ to be very
similar to the oxides, with O replaced by halogen and the divalent
cation with monovalent Cs, also resulting in divalent
Tl formally with one electron in a $6s$ orbital, similar
to the tetravalent Bi-based problem.
It remains unknown, however,
whether hole doping can make the thallium halides superconduct\cite{Retuerto2}.
While we note an overall similarity with
the electronic structure of the previously
discussed $AB$O$_{3}$ compounds,
the $\text{Tl}\textendash 6s - \text{halogen}\textendash a_{1g}$
band splitting is considerably smaller.
Also, even though the difference between the CsTlF$_{3}$ and CsTlCl$_{3}$
lattice constants is quite significant (Table~\ref{tab1}),
we see only a slight change in the band-widths of their
$\text{Tl}\textendash 6s - \text{halogen}\textendash a_{1g}$
band manifolds.

We proceed now to quantifying the observed differences
in the electronic structures of the $ABX_{3}$
compounds in terms of the
$B\textendash s - X\textendash p$ hybridization strength
$t_{sp\sigma}$ and the charge-transfer energy $\Delta$.
As was shown in Ref.~\onlinecite{Khazraie2}, 
these are the two main parameters that determine the character of
the empty states crossing the Fermi level.
We should emphasize that in our
convention  $\Delta$ is defined as the difference between
the on-site energies of the $B\textendash s$ atomic orbital
and of the $X\textendash a_{1g}$ MO:
$\Delta = \epsilon(B\textendash s) - \epsilon(X\textendash a_{1g})$\cite{Khazraie2}.
$t_{sp\sigma}$ and  $\Delta$ are obtained
by calculating MLWF-based TB models for our systems.
Similarly to Ref.~\onlinecite{Khazraie}, here we 
consider 10 orbitals per formula unit:
one $B\textendash s$
and three $X\textendash p$ orbitals per
each of the three anions in a simple cubic unit cell. 
As one can see in Figs.~\ref{fig2}~(a)-(e) and \ref{fig4}~(a)-(b),
the resulting TB models have band dispersions that agree
well with the GGA band structures.
However, small deviation exists
at the $R$ point of MgPO$_3$, CaAsO$_3$, and SrSbO$_3$, due to hybridization
with higher energy states, as discussed earlier in the paper.
We have chosen not to consider the
higher energy $B\textendash p$ orbitals in the TB models
because our main focus is on the low-energy scale electron
removal and addition states relevant for the physical properties
at relatively low temperatures. 

The hopping integrals $t_{sp\sigma}$ are given in Table~\ref{tab1}.
While for the $AB$O$_{3}$ series the value of $t_{sp\sigma}$ decreases
rather monotonically from MgPO$_{3}$ to RaMcO$_{3}$ and is inversely
related to the unit cell's size, the $t_{sp\sigma}$ 
values
of the two halides
are surprisingly similar given the big difference between their unit cells' sizes.
In order to obtain the charge-transfer energy
$\Delta = \epsilon(B\textendash s) - \epsilon(X\textendash a_{1g})$,
we have applied a basis set transformation
from oxygen atomic to oxygen molecular orbitals using the table in
Fig.~\ref{fig1}~(b).
The resulting charge-transfer energy values $\Delta$ are listed in Table~\ref{tab1}.
In the $AB$O$_{3}$ series, $\Delta$ varies widely from a positive
value of 1.83~eV in MgPO$_{3}$ to a negative value of -8.86~eV in RaMcO$_{3}$,
thus marking a difference between MgPO$_{3}$ and CsTlF$_3$,
as being
\emph{positive charge-transfer energy} compounds, and
the rest of the $AB$O$_3$ systems and CsTlCl$_3$, as being
\emph{negative charge-transfer energy} compounds.
We note again that in all cases, except perhaps RaMcO$_3$,
the total bonding\textendash anti-bonding splitting strongly dominates
over the charge transfer energy, making the latter less important
than the hopping integrals.

The dramatic decrease of $\Delta$
in the $AB$O$_{3}$ series with $A$ and $B$ moving down
the periodic table can be understood in terms of relativistic lowering of
the $6s$ and $7s$ orbital energies in the heavy elements Bi and Mc\cite{Pyykko}.
In RaMcO$_{3}$, it takes an extreme form, which, combined with the reduced $t_{sp\sigma}$
hybridization, results in an almost ionic character of the Mc$^{3+}$ ionization state,
in agreement with earlier studies \cite{Keller}.
This can be clearly seen from the RaMcO$_{3}$ projected DOS
shown in Fig.~\ref{fig3}~(e).
Similarly to Ba(Sr)BiO$_{3}$,
RaMcO$_{3}$ is therefore expected to bond-disproportionate
in its ground state as 
$2\text{Mc}^{3+}\underline{L}^{2}\rightarrow
[\text{Mc}^{3+}]_{\text{large}}+[\text{Mc}^{3+}\underline{L}^{2}]_{\text{small}}$,
with most of the action happening on oxygens.
On the other hand, also interesting is the fact that BaSbO$_{3}$
can become superconducting upon substituting Sb with Pb but not with Sn\cite{Cava2}.
This again was explained in terms of the relativistic energy
lowering of Pb$\textendash 6s$ states with respect to
Sn$\textendash 5s$ states\cite{Singh}.
The maximal $T_{\text c}$ observed in BaPb$_{x}$Sb$_{1-x}$O$_{3}$ is 3.5 K,
which is unexpectedly low compared with that of the bismuthates.
At this point, we can only speculate that if there is a relation
between $T_{\text c}$ and how negative $\Delta$ is,
then optimally hole-doped RaMcO$_{3}$
might have a very high $T_{\text c}$.
Such a relation can be rooted
in electron-phonon coupling which might become most efficient
in the presence of ligand holes, but of course more theoretical
and experimental studies are required to give some validity to this idea.
We should also note that the density
of states at $E_{\text F}$ is very high
in RaMcO$_3$ because of the narrowing
of the anti-bonding band and this could also support a higher $T_{\text c}$.

\begin{figure}[tb]
\begin{center}
\includegraphics[width=0.49\textwidth]{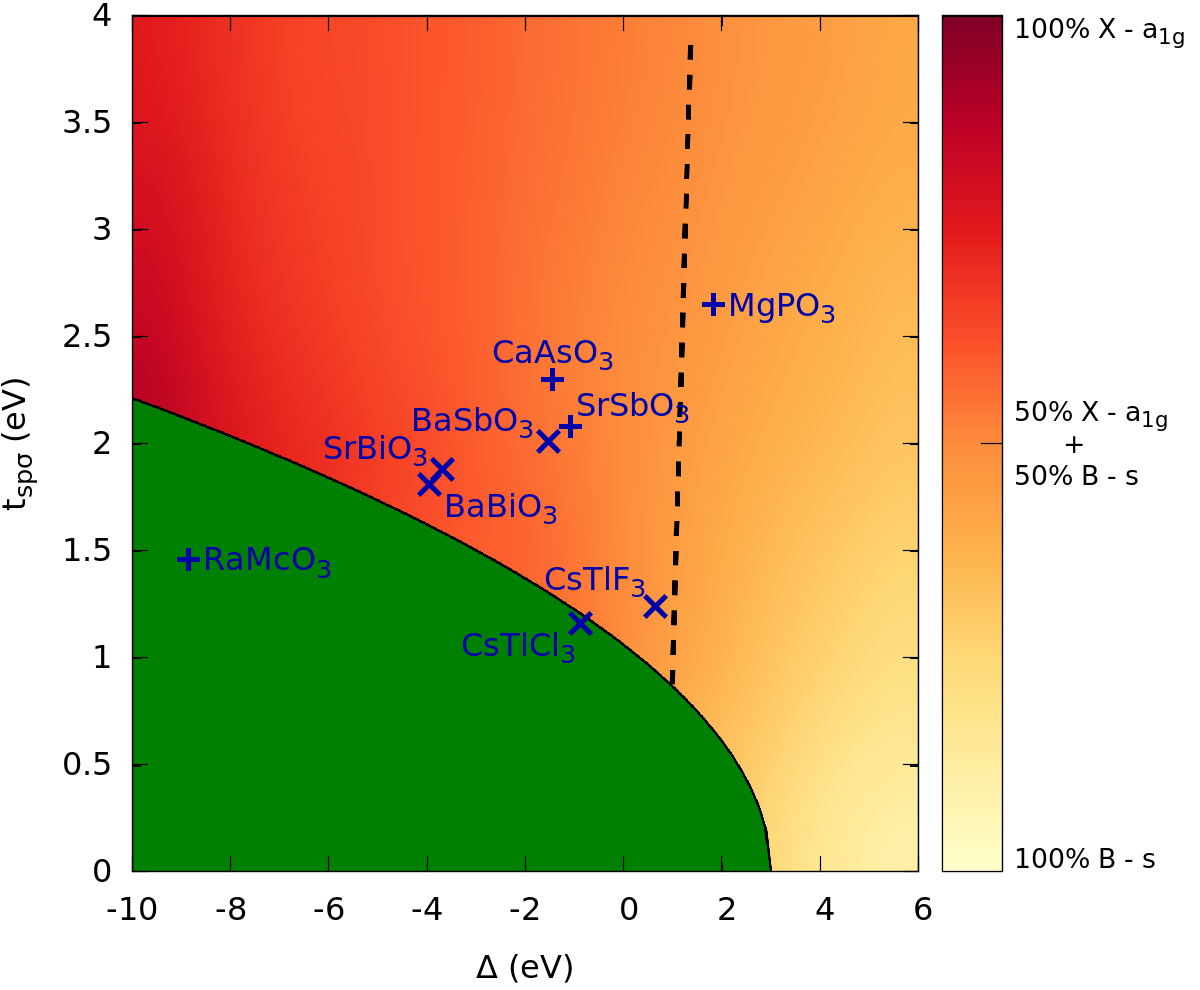} 
\caption{
The phase diagram of Ref.~\onlinecite{Khazraie2}
representing the dominant character of holes as a function
of the charge-transfer energy $\Delta$
and hybridization $t_{sp\sigma}$.
Symbols $+$ and $\times$ mark the parameters relevant
to the hypothetical and existing
cubic $s\textendash p$ $ABX_3$ perovskites, respectively.
Green, yellow, and red colors
represent the amount of the $X\textendash e_{g}$,
$X\textendash a_{1g}$, and $B$ cation $s$ orbital
contributions to the holes' character, respectively.
}
\label{fig5}
\end{center}
\end{figure}

Finally, with the hopping integrals $t_{sp\sigma}$ and
charge-transfer energies $\Delta$
of our studied $ABX_{3}$ compounds at hand, we can mark their positions
on the phase diagram proposed in Ref.~\onlinecite{Khazraie2}.
There will be some degree of approximation involved because
this phase diagram was obtained for BaBiO$_{3}$
in a bond-disproportionated state using
the value of the oxygen band-width $W$ specific to BaBiO$_{3}$,
but this is not going to obscure observation of general
trends that we are most interested in.
As explained in Ref.~\onlinecite{Khazraie2},
the colors in this phase diagram represent the dominant character
of the empty (hole) states above the Fermi level as a function
of hybridization and charge-transfer energy.
While there is a sharp boundary around the green region where holes
reside on $X\textendash e_g$ orbitals, the $B\textendash s$ (yellow)
and $X\textendash a_{1g}$ (red)
orbitals are always mixed by hybridization and therefore there
is a gradual crossover between the yellow and red regions.
The black dashed line marks equal contributions from the $B\textendash s$
and $X\textendash a_{1g}$ orbitals to the holes' character.
Figure~\ref{fig5}
is showing now that SrBiO$_{3}$ and BaBiO$_{3}$ are relatively
deep in the $X\textendash a_{1g}$ region, while
MgPO$_3$, CaAsO$_{3}$, SrSbO$_{3}$, BaSbO$_3$, and CsTlF$_{3}$
are close to having equal $B\textendash s$
and $X\textendash a_{1g}$ orbital contributions.
As for RaMcO$_{3}$ and CsTlCl$_{3}$,
they land in the O$\textendash e_g$ region but from their projected
DOS [Fig.~\ref{fig3}~(e) and Fig.~\ref{fig4}~(d)]
we know that their hole character is strongly O$\textendash a_{1g}$.
This discrepancy
is due to the mentioned approximations,
but it is obvious that RaMcO$_{3}$ must in any case
be located very far inside
the O$\textendash a_{1g}$ region close to the O$\textendash e_g$ border.

\section{Conclusions}

In this paper,
we have used {\it ab initio} methods to study the electronic structures
of the following $s\textendash p$ cubic perovskites $ABX_{3}$:
the experimentally available BaSbO$_3$, SrBiO$_{3}$, BaBiO$_{3}$, CsTlF$_{3}$,
and CsTlCl$_{3}$, as well as the hypothetical MgPO$_{3}$, CaAsO$_{3}$,
SrSbO$_{3}$, and RaMcO$_3$. We have used Wannier functions based tight-binding modeling 
to calculate their hybridization strengths $t_{sp\sigma}$
between the $B\textendash s$ and $X\textendash p$ atomic orbitals and charge-transfer
energies $\Delta$, which are the two most important parameters
that determine the nature of the systems' holes.
These calculations have elucidated several
trends in $t_{sp\sigma}$ and $\Delta$ as one moves across the periodic table,
such as the relativistic energy lowering of the $B\textendash s$ orbital in heavy
$B$ cations leading to strongly negative $\Delta$ values.
Our results have been discussed in connection with the general
phase diagram for $s\textendash p$ cubic perovskites proposed
in Ref.~\onlinecite{Khazraie2}.

Also, some considerations have been offered regarding a possible relation
between the highest achievable
superconducting transition temperatures and
certain features of the systems' electronic structures, such as
the charge-transfer energy $\Delta$ and the interatomic orbital
hybridization $B\textendash s - {\text O}\textendash 2p_{\sigma}$
$t_{sp\sigma}$,
with the latter primarily controlling the conduction band width
and the DOS at the Fermi level.
In particular, we observed that
the more negative the $\Delta$ value is
the more of the $X\textendash a_{1g}$ character will be there
in the valence and conduction
bands provided that the hybridization strength is large enough
to still push out the $X\textendash a_{1g}$ bound state.
In the case of RaMcO$_3$ and also Ba(Sr)BiO$_3$,
$\Delta$
is very negative and so the amount
of the $X\textendash a_{1g}$ character in the anti-bonding state will be
the largest.
As $\Delta$ becomes even more negative
there will be a point where the $X\textendash a_{1g}$ state no
longer is pushed above the $X\textendash e_{g}$ state,
as illustrated in the phase diagram
of Fig.~\ref{fig5}.
Given the significantly
larger maximal $T_{\text c}$
in the hole-doped bismuthates Ba(Sr)BiO$_3$ than in the hole-doped antimonates
BaSbO$_3$,
superconductivity is apparently enhanced if the charge character
of the hole states is mostly $X\textendash a_{1g}$,
which can possibly be traced down to the enhanced coupling
with the breathing mode phonon.
However, an even more important
feature of the electronic structure
that can be responsible for an enhanced $T_{\text{c}}$
is the increase of the DOS at the Fermi level in BaBiO$_3$,
due to its reduced $t_{sp\sigma}$ value.
For these reasons,
we expect that optimally hole-doped RaMcO$_3$ might have a very high $T_{\text{c}}$.


\section*{Acknowledgments}
This work was supported by Natural Sciences and
Engineering Research Council (NSERC) for Canada, CIfAR,
and the Max Planck - UBC Stewart Blusson Quantum Matter
Institute.

\end{document}